\documentclass[aps,prd,twocolumn,groupedaddress,superscriptaddress]{revtex4-2}

\usepackage{lipsum}
\usepackage{graphicx}  % needed for figures
\usepackage{dcolumn}   % needed for some tables
\usepackage{bm}        % for math
\usepackage{amssymb}   % for math
\usepackage{amsmath}
\usepackage{xcolor}
\usepackage{amsmath}
\usepackage{amsfonts}
\usepackage{hyperref}
\usepackage{float}
\usepackage{natbib}

\hypersetup{colorlinks=true, linkcolor=blue, citecolor=blue, urlcolor=blue}

\begin{document}

\preprint{APS/123-QED}
\title{Exploring Transport Properties of Quark-Gluon Plasma in Flavor-Dependent Systems with a Holographic Model}
\author{Bing Chen}\affiliation{School of Nuclear Science and Technology, University of South China,
 Hengyang 421001, People's Republic of China}
\author{Xun Chen}
\email{chenxun@usc.edu.cn}\affiliation{School of Nuclear Science and Technology, University of South China,
 Hengyang 421001, People's Republic of China}
\affiliation{Key Laboratory of Quark and Lepton Physics (MOE), Central China Normal University, Wuhan 430079,China}
\affiliation{Key Laboratory of Advanced Nuclear Energy Design and Safety, Ministry of Education, Hengyang, 421001, China}
\author{Xiaohua Li}
\email{lixiaohuaphysics@126.com}\affiliation{School of Nuclear Science and Technology, University of South China, Hengyang 421001, People's Republic of China}
\affiliation{Key Laboratory of Advanced Nuclear Energy Design and Safety, Ministry of Education, Hengyang, 421001, China}
\author{Zhou-Run Zhu}
\email{zhuzhourun@zknu.edu.cn}\affiliation{School of Physics and Telecommunications Engineering,
 Zhoukou Normal University, Zhoukou 466001, China}
\author{Kai Zhou}
\email{zhoukai@cuhk.edu.cn}\affiliation{School of Science and Engineering, The Chinese University of Hong Kong, Shenzhen (CUHK-Shenzhen), Guangdong, 518172, China}
\affiliation{Frankfurt Institute for Advanced Studies, Ruth Moufang Strasse 1, D-60438, Frankfurt am Main, Germany}

\begin{abstract}
Based on the holographic model, which incorporates the equation of state (EoS) and baryon number susceptibility for different flavors, we calculate the drag force, jet quenching parameter, and diffusion coefficient of the heavy quark at finite temperature and chemical potential. The holographic results for the diffusion coefficient align with lattice data for $N_f = 0$ and $N_f = 2+1$, falling within their error margins. The holographic diffusion coefficient for heavy quark in the systems of different flavors is compatible with estimates from ALICE data. The jet quenching parameter in our model
 demonstrates strong consistency with the estimations obtained from Bayesian analysis of data from both RHIC and LHC for different flavors. We can confirm the model provides a good description of the transport properties of QGP. The work reinforces the potential of bottom-up holographic model in advancing our understanding of transport properties of hot and dense quark-gluon plasma.
\end{abstract}
\maketitle

\section{Introduction}
\label{sec-int}
The Quark-Gluon Plasma (QGP), created in high-energy nuclear collision experiments such as those at the Relativistic Heavy Ion Collider (RHIC) and the Large Hadron Collider (LHC), is a type of extreme Quantum Chromodynamics (QCD) matter that exhibits strong coupling and is influenced heavily by non-perturbative phenomena. Perturbative QCD proves inadequate in these strong coupling scenarios \cite{Baier:1996kr,Eskola:2004cr}, hence lattice QCD methods are employed to explore the static equilibrium properties of this state of matter. In addition, there exists an alternative non-perturbative approach known as the AdS/CFT correspondence \cite{Maldacena:1997re,Witten:1998qj,Casalderrey-Solana:2011dxg}, which facilitates the investigation of dynamic properties of QGP in a strong coupling regime. This correspondence draws a parallel between $\mathcal{N}=4$ SU($\rm N_c$) super-Yang-Mills theory and type IIB string theory on a combined $\rm AdS_5 \times S^5$ space, offering a robust method to analyze strongly interacting gauge theories when the number of color charges $\rm N_c$ is large, and the 't Hooft coupling is also substantial. The original form of this duality linked an asymptotically AdS space to a conformal gauge theory at a temperature of absolute zero. However, recognizing that the properties of QGP are dependent on temperature, researchers have strived to broaden this duality to encapsulate holographic models that depict the QGP at non-zero temperatures. This extension has been explored through both top-down \cite{Polchinski:2000uf,Sakai:2004cn} and bottom-up \cite{He:2013qq,Yang:2015aia,Gursoy:2007er,Alho:2012mh,Panero:2009tv,Dudal:2015kza,Dudal:2015wfn,Zollner:2024iza,Cao:2024jgt,Wang:2024rim,Chen:2024mmd,
Cai:2024eqa,Ahn:2024jkk,Ahn:2024gjf,Jokela:2024xgz,Bea:2024xgv} approaches in various studies.

In the strong-coupling limit of the gauge theory effectively reduces to classical gravity on $AdS_5$ \cite{Maldacena:1997re}. For practical purposes, the most useful feature of the correspondence is that it maps the perturbatively inaccessible strong-coupling regime of a quantum field theory (QFT) to a weakly interacting gravity theory. Gauge/gravity duality by contrast drastically simplifies such calculations at strong coupling. For instance, it reduces the calculation of real-time correlation functions to solving linear wave equations in classical gravity. These simplifications, however, come at a price: QFTs with a holographic gravity dual are very particular and in fact rather different from realistic theories such as QCD. Due to these restrictions, holographic gauge theories with a gravity dual can at best serve as useful toy models for strongly coupled real-world systems.\\

The gauge/gravity duality has proven its worth in several key respects, as outlined by \cite{Probst:2017vsq}. Firstly, it highlights that the low-energy characteristics of physical systems are often independent of the specific ultraviolet (UV) degrees of freedom. Instead, these universal traits are governed by infrared (IR) physics, being shared among various theories that converge near similar IR fixed points in their renormalization group trajectories. Identifying such universal properties across a broad spectrum of holographic theories suggests that these may be intrinsic features of strong coupling, potentially extending to real-world, non-holographic systems. A prime example is the ratio of shear viscosity \(\eta\) to entropy density \(s\), which, when normalized by \(\hbar/k_B\), consistently yields \(1/4\pi\) across many holographic systems \cite{Casalderrey-Solana:2011dxg}. This ratio significantly deviates from values in weakly coupled theories and intriguingly aligns with experimental findings from quark-gluon plasma. Secondly, the duality serves as an invaluable tool in the absence of alternative methods for computing properties within strongly interacting systems. The gravitational dual provides a geometric framework to articulate the complexities of strongly coupled dynamics, which are not satisfactorily captured by the traditional quasi-particle approach of weakly interacting systems \cite{Hartnoll:2016apf}. Thirdly, the novel insights gleaned from the gauge/gravity duality have spurred advancements in field theory and gravity, yielding results with standalone significance beyond their holographic context. Notable among these are the uncovering of novel transport coefficients in hydrodynamics \cite{Baier:2007ix,Bhattacharyya:2007vjd} and the revelation of exceptions to the no-hair theorems pertaining to black holes \cite{Gubser:2008px}.

An intriguing aspect of QCD is the confinement-deconfinement phase transition, characterized by a substantial increase in the QCD coupling constant. In proximity to the transition temperature, there's a notable suppression of quarkonium states because of the nature of the deconfinement occured \cite{PHENIX:2006gsi,ALICE:2013osk,Matsui:1986dk,Kaczmarek:2005ui,Hashimoto:2014fha}. These phenomena can also be probed by employing the AdS/CFT correspondence and developing various specialized holographic dual models \cite{Maldacena:1998im,Rey:1998ik,BitaghsirFadafan:2015zjc,Iatrakis:2015sua,Chen:2017lsf}. Within the array of holographic models tailored to QCD, the construction rooted in the Einstein-Maxwell-dilaton(EMD) gravity framework stands out \cite{DeWolfe:2010he,Chen:2022goa,Rougemont:2023gfz,Jarvinen:2022doa,Knaute:2017opk,Arefeva:2024vom,Dudal:2018ztm,Dudal:2017max,Fu:2024wkn}. In early papers \cite{Gubser:2008yx,Gubser:2008ny}, the authors tried to find a five-dimensional gravitational theory that has black hole solutions mimics QCD property with a ansatz potential of dilaton field. These papers demonstrate that the bottom-up holographic models can capture the property of sound of speed and shear viscosity for QGP. In Ref. \cite{DeWolfe:2010he}, the authors constructed a holographic model which can describe the entropy and quark susceptibility of lattice results well. Then, the model is used to give a prediction of CEP.

In the recent works \cite{Chen:2024ckb,Chen:2024mmd}, we developed a machine learning assisted EMD model which includes the information of equation of state and baryon number susceptibility from lattice QCD. In our works, we used machine-learning techniques, specifically automatic differentiation, to determine the six parameters within our holographic setup. This approach allowed us to efficiently explore a complex parameter space and find optimal solutions that best reproduce QCD thermodynamic data across a wide temperature range. The parameter space is high-dimensional and non-linear, making traditional fitting methods less effective. Beside, machine-learning algorithms can handle complex constraints and multiple optimization objectives simultaneously. Our prediction for the position of CEP is close to other theoretical models, which shows the validity of our model. Further, we calculated the heavy quark potential in our model \cite{Guo:2024qiq}. The results of heavy quark potential is in agreement with new lattice results for $N_f = 2 + 1$, which is again confirmed that our model captures the property of QCD.

Quark jets traversing the QGP represent one of the most intriguing phenomena generated in high-energy nuclear collisions, and their interaction with the surrounding medium presents a multifaceted challenge in contemporary physics. A fundamental experimental measure tied to the study of quark jets is the transport coefficient, commonly known as the jet quenching parameter. This parameter provides crucial insights into the quark energy dissipation within the hot and dense environment created in experiments at RHIC and LHC \cite{STAR:2005gfr,Yin:2013zea,ATLAS:2010isq,JET:2013cls,CMS:2011iwn}.
The jet quenching parameter is quantified as the mean square of the transverse momentum imparted from a parton to the surrounding medium per unit of the parton's mean free path \cite{DEramo:2010wup}. Numerous models have been used to calculate the jet quenching parameter \cite{Wang:1992qdg,Liu:2006ug,Renk:2006sx,Jiang:2022uoe,Jiang:2022vxe}. Furthermore, the energy depletion of partons as they pass through the QGP can be examined via the drag force exerted on heavy quarks moving within the plasma \cite{Rougemont:2015wca}. Given the intensity of the interactions involved, holographic QCD models have become invaluable in probing the dynamics underlying these processes. They offer substantial insights into the characteristics of both the jet streams and the QGP, along with the intrinsic interactions that culminate in the loss of quark energy.

\begin{figure}
    \centering
    \includegraphics[width=\columnwidth]{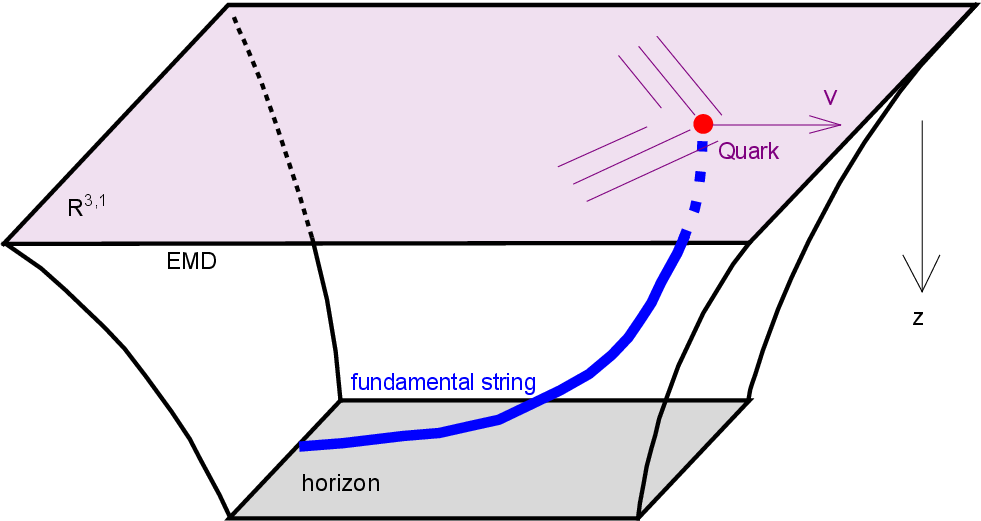}
    \caption{\label{n} A sketch of holographic drag force. $z$ is the fifth dimension.
}
\end{figure}

Within the framework of the AdS/CFT correspondence, a heavy quark is represented by a fundamental string anchored to a flavor brane. The endpoint of the string is perceived as the quark in the boundary field theory, while the string itself symbolizes the gluonic field enveloping the quark as in Fig. \ref{n}. The resistance experienced by a moving quark in the plasma is reflected by the momentum flux from the trailing end of an open string into the deeper AdS space \cite{Rougemont:2015wca,Mes:2020vgy,Peng:2024zvf,Gubser:2006qh,Caceres:2006dj,Cheng:2014fza,Gubser:2006bz,Zhu:2019ujc,Domurcukgul:2021qfe,Grefa:2022sav,Giataganas:2013hwa,Zhang:2018mqt,Zhu:2020wds,Andreev:2017bvr,Chen:2023yug,Mykhaylova:2020pfk,Gubser:1996de,Gubser:2007zz,Gubser:2009fc,Gursoy:2009kk,Zhou:2022izh}.
In the vein of this duality, the jet quenching parameter is tied to the thermal expectation of the Wilson loop operator, which is light-like and constructed from the trajectories of the string's endpoints. Numerous efforts have concentrated on calculating this parameter within the scope of the AdS/CFT duality, providing valuable insights into the intricate dynamics governing quark propagation and energy loss in the QGP \cite{Rougemont:2015wca,Zhou:2022izh,Zhou:2024oeg,Buchel:2006bv,Herzog:2006gh,Li:2014dsa,Grefa:2023hmf,Apolinario:2022vzg,Sadeghi:2013dga,Wang:2016noh,Du:2023qst,Horowitz:2015dta,Heshmatian:2018wlv,BitaghsirFadafan:2017tci,Li:2014hja,Zhang:2024ebf,Zhang:2023kzf,Hou:2021own,Xing:2021bsc,Li:2020kax,Critelli:2017oub}.

Employing the gauge/gravity duality as our investigative tool, we utilize the dynamical holographic QCD model outlined in Ref. \cite{Chen:2024ckb}. This model proves particularly apt for our analysis as it encapsulates the effects of temperature across both confined and deconfined phases of QCD, while also incorporating the chemical potential variations. Moreover, it faithfully represents the equation of state and baryon number susceptibility observed in lattice QCD. Given these comprehensive attributes, the model stands as an optimal tool for probing the intricate characteristics of the QGP. In this study, we also aim to examine whether the model's predictions align with estimates derived from lattice QCD, perturbative QCD, etc, and whether the model provides an good description of the transport properties of QGP.

The structure of the article is organized as follows: Sec. \ref{1} provides a brief review of the holographic QCD model established by the EMD gravity introduced in \cite{Chen:2024ckb}. Sec. \ref{2} discusses the drag force experienced by a heavy quark in motion within the holographic QCD dynamics model. In Sec. \ref{3}, we calculate the diffusion coefficient of the heavy quark. Sec. \ref{4} is devoted to the examination of the jet quenching parameter. Sec. \ref{5} presents the overall summary and conclusions of the article.

\section{The EMD framework}
\label{1}
First, we review the five-dimensional EMD system of our model \cite{Chen:2024ckb,Chen:2024mmd}. This system comprises a gravitational field $g_{\mu\nu}$, a Maxwell field $A_{\mu}$, and a dilaton field $\phi$. In the Einstein frame, its action is expressed by the following equation:
\begin{equation}\begin{aligned}
&S_E = \frac{1}{16\pi G_5} \\&\int d^5x \sqrt{-g} \left[ R - \frac{f(\phi)}{4} F^2 - \frac{1}{2} \partial_{\mu}\phi\partial^{\mu}\phi - V(\phi) \right].\end{aligned}\label{eq1}
\end{equation}
Here, $R$ is the Ricci scalar, $F_{\mu\nu} = \partial_{\mu}A_{\nu} - \partial_{\nu}A_{\mu}$ is the electromagnetic field tensor, with $f(\phi)$ being the gauge kinetic function coupling to the gauge field $A_{\mu}$, $F$ is the Maxwell field tensor, $V(\phi)$ is the dilaton potential, and $G_5$ is the five-dimensional Newton constant. The explicit forms of the gauge kinetic function $f(\phi)$ and the dilaton potential $V(\phi)$ can be consistently solved through the equations of motion.

We propose the following metric ansatz
\begin{equation}
ds^2 = \frac{L^2 e^{2A(z)}}{z^2}\left[-g(z)dt^2 + \frac{dz^2}{g(z)} + d\vec{x}^2\right],\label{eq2}
\end{equation}
where $z$ is the holographic radial coordinate in the fifth dimension and the AdS$_5$ space radius $L$ is set to $L=1 \, \rm GeV^{-1}$ by comparing with the equation of state from lattice QCD \cite{Chen:2024ckb,Chen:2024mmd}.

To obtain analytical solutions, we assume the forms of $f(\phi)$ and $A(z)$ along with some parameters. We adopt the metric ansatz
\begin{equation}
A(z) = d \ln(az^2 + 1) + d \ln(bz^4 + 1),\label{eq3}
\end{equation}
and the form of the gauge kinetic function $f(z)$ as
\begin{equation}
f(z) = e^{cz^2 - A(z) + k}.\label{eq4}
\end{equation}
$A(z)$ is set to mimic the correct behavior of entropy and to constrain the temperature-dependent model. The function $f(z)$ describes the model's dependence on the chemical potential, which is fixed by the baryon number susceptibility. Then, we can derive
\begin{equation}\begin{aligned}
&g(z) = 1 - \frac{1}{\int_0^{z_h} dx\, x^3 e^{-3A(x)}} \\
&\times\left[\int_0^{z} dx\, x^3 e^{-3A(x)} + \frac{2c\mu^2 e^k}{\left(1 - e^{-cz^2_h}\right)^2} det G\right],\end{aligned} \label{eq5}
\end{equation}
\begin{equation}
\phi'(z) = \sqrt{6 \left(A'^2 - A'' - \frac{2A'}{z}\right)}, \label{eq6}
\end{equation}
\begin{equation}
A_t(z) = \frac{\mu \left(e^{-cz^2} - e^{-cz^2_h}\right)}{1 - e^{-cz^2_h}},\label{eq7}
\end{equation}
and the dilaton potential as
\begin{equation}\begin{aligned}
&V(z) = -\frac{3 z^2 ge^{-2A}}{L^2} \\&\times\left[ A'' + A' \left(3A' - \frac{6}{z} + \frac{3g'}{2g}\right) - \frac{1}{z} \left(- \frac{4}{z} + \frac{3g'}{2g}\right) + \frac{g''}{6g} \right].\end{aligned}\label{eq8}
\end{equation}
The determinant $G$ is given by
\[
\det G = \begin{vmatrix}
\int_0^{z_h} dy\, y^3 e^{-3A(y)} & \int_0^{z_h} dy\, y^3 e^{-3A(y)-cy^2} \\
\int_{z_h}^z dy\, y^3 e^{-3A(y)} & \int_{z_h}^z dy\, y^3 e^{-3A(y)-cy^2}
\end{vmatrix}.
\]
The Hawking temperature and entropy of this black hole solution are given by the following formulas,
\begin{equation}\begin{aligned}
&T = \frac{z_h^3 e^{-3A(z_h)}}{4\pi\int_0^{z_h} dy\, y^3 e^{-3A(y)}} [ 1 + \\&\frac{2c\mu^2 e^k \left( e^{-cz_h^2}\int_0^{z_h} dy\, y^3 e^{-3A(y)} - \int_0^{z_h} dy\, y^3 e^{-3A(y) -cy^2}\right)}{(1 - e^{-cz_h^2})^2}],\end{aligned}\label{eq9}
\end{equation}
\begin{equation}
s = \frac{e^{3A(z_h)}}{4G_5 z_h^3}.\label{eq10}
\end{equation}
For convenient study of our holographic probes of interest, we use the metric in the string frame :
\begin{equation}
ds_s^2 = \frac{L^2 e^{2A_s(z)}}{z^2} \left( -g(z)dt^2 + \frac{dz^2}{g(z)} + dx_1^2 + dx_2^2 + dx_3^2 \right),\label{eq11}
\end{equation}
where $A_s(z) = A(z) + \sqrt{\frac{1}{6}}\phi(z)$.

There are three undetermined parameters $a$, $b$, and $d$ in $A(z)$, and two parameters $c$ and $k$ in $f(z)$, along with the Newton constant $G_5$, making in total a six-dimensional parameter space. The six parameters, as determined by machine learning from the data of the EoS and baryon number susceptibility from the lattice QCD, vary among pure gluon, 2-flavor, and 2+1-flavor systems, as shown in Table \ref{table:parameter} \cite{Chen:2024ckb}. The QCD phase transition is believed to exhibit strong coupling around the phase transition temperature \( T_c \). Our estimation utilizes lattice QCD simulations. The fitting range for our model is approximately from \( 0.9 T_c \) to \( 2 T_c \). We employ extrapolation for temperatures beyond this range. Thus, we believe the results are reliable from \( 0.9 T_c \) to \( 2 T_c \).
\begin{table}[htbp]
	\centering
	\begin{tabular}{|c|c|c|c|c|c|c|c|}
		\hline
		&$a$&$b$&$c$&$d$&$k$&$G_5$&$T_c$ \\
		\hline
		$N_f = 0$&0&0.072&0&-0.584&0&1.326&0.265  \\
        \hline
		$N_f = 2$&0.067&0.023&-0.377&-0.382&0&0.885&0.189  \\
        \hline
        $N_f = 2+1$&0.204&0.013&-0.264&-0.173&-0.824&0.400&0.128   \\
        \hline

	\end{tabular}
\caption{Parameters given by the machine learning of pure gluon system, 2-flavor, and 2+1-flavor system, respectively. $T_c$ is the critical temperature calculated by $c_s^2$ inflection. The unit of $T$ is GeV. The unit of $G_5$ is $\rm GeV^3$. The units of $a$ and $c$ are $\rm GeV^2$. The unit of $b$ is $\rm GeV^4$.}
\label{table:parameter}
\end{table}

\section{Drag Force}
\label{2}
In this section, we calculate the drag force in three different systems. Firstly, we consider a heavy quark moving with a constant velocity along one spatial direction, denoted by $x$. Therefore, with the static gauge $\tau = t$, $\sigma = z$, the embedding function of the heavy quark string is $X = \{t, x(t, z), z\}$. The action of the fundamental string is given by the Nambu-Goto action:
\begin{equation}\begin{aligned}
&S = -\frac{1}{2\pi\alpha'} \int d\sigma d\tau \sqrt{-\gamma} \\&= -\frac{1}{2\pi\alpha'} \int d\sigma d\tau \sqrt{-G_{tt}G_{zz} - G_{tt}G_{xx}x'^2 - G_{xx}G_{zz}\dot{x}^2},\end{aligned}\label{eq12}
\end{equation}
where dots and primes denote derivatives with respect to $\tau$ and $\sigma$, respectively, and $G$ stands for the components of the background metric. The only one new parameter $\alpha'$ is taken to be 2.5 in this paper. To study the dynamics of the string, we use the metric from Eq. (\ref{eq11}). In the static gauge, the corresponding Lagrangian density takes the form,
\begin{equation}
\mathcal{L} = -\frac{1}{2\pi\alpha'} \frac{L^2 e^{2A_s(z)}}{z^2} \sqrt{1 - \frac{\dot{x}^2}{g(z)} + g(z)x'^2(z)}.\label{eq13}
\end{equation}
The equation of motion for $x$ is then,
\begin{equation}
\partial_t \left( \frac{\dot{x}}{\sqrt{-\gamma'}} \right) - \frac{z^2 g(z)}{L^2 e^{2A_s(z)}}\partial_z \left( \frac{L^2 e^{2A_s(z)} g(z) x'}{z^2\sqrt{-\gamma'} } \right) = 0.\label{eq14}
\end{equation}
A static string stretching from the boundary to the horizon, $x(t, z) = \text{Constant}$, is a trivial solution to this equation. For a string with an endpoint moving at a constant velocity $v$ on the boundary, we choose the following trial solution,
\begin{equation}
X = \{t, vt + \xi(z), z\},\label{eq15}
\end{equation}
the equation of motion can be written as
\begin{equation}
\frac{L^2 e^{2A_s(z)}g(z)\xi'(z)}{\sqrt{-\gamma}z^2} = \text{const} \equiv \pi_x,\label{eq16}
\end{equation}
where $\pi_x$ is the conserved quantity on the worldsheet. The equation for $\xi$ is obtained as follows,
\begin{equation}
\xi'(z) = \pm \frac{\pi_x}{g(z)}\sqrt{ \frac{g(z) - v^2}{\frac{L^4 e^{4A_s(z)}g(z)}{z^4} - \pi_x^2}}. \label{eq17}
\end{equation}
The requirement that the function under the square root be real determines the constants of motion,
\begin{equation}
\begin{aligned}
& g(z_s) - v^2 = 0 \\
& \frac{L^4 e^{4A_s(z_s)}g(z_s)}{z_s^4} - \pi_x^2 = 0 \\
& \Rightarrow \quad \pi_x = \frac{L^2 e^{2A_s(z_s)}v}{z_s^2}
\end{aligned}
\label{eq18}
\end{equation}

Finally, by substituting Eq. (\ref{eq18}) into Eq. (\ref{eq17}), we can solve for the string solution,
\begin{equation}
\xi'(z) = -\frac{L^2 e^{2A_s(z_s)}v}{z_s^2} \frac{1}{g(z)}\sqrt{\frac{g(z) - g(z_s)}{\frac{L^4 e^{4A_s(z)}g(z)}{z^4} - \frac{L^4 e^{4A_s(z_s)}g(z_s)}{z_s^4}}}.\label{eq19}
\end{equation}
$\pi^1_t$ and $\pi^1_x$ represent the flow of energy and momentum along the string, respectively. They are given by
\begin{equation}
\begin{pmatrix}
\pi^1_t \\
\pi^1_x \\
\end{pmatrix}
= \frac{T_0 L^4}{\sqrt{-\gamma}}\frac{ e^{4A_s(z)}}{z^4}
\begin{pmatrix}
-v g(z) \xi'(z) \\
g(z) \xi'(z) \\
\end{pmatrix}.\label{eq20}
\end{equation}
It is straightforward to show that in the equation of motion, $\pi_x^1$ is indeed a constant, the same as $\pi_x$ in Eq. (\ref{eq17}). Similarly, like in the case of the $\mathcal{N}=4$ SYM plasma, we have $\pi_t^1 = -v \pi_x^1$. This means that if we drag the quark at a constant velocity, the fraction of energy flow $\pi_t^1$ at a given point on the string is constant. This is the energy dissipation of the quark into the surrounding medium. Thus, the drag force can be obtained as follows \cite{Rougemont:2015wca},
\begin{equation}
F_{\text{drag}} =\frac{dp}{dt}=\frac{dE}{dx}= -\pi_x^1= -\frac{1}{2\pi\alpha'} \frac{L^2e^{2A_s(z_s)}v}{ z_s^2}.\label{eq21}
\end{equation}
The drag force in the AdS/Schwarzchild background can be obtained as follows \cite{Gubser:2006qh},
\begin{equation}
F_{\text{drag}}^{\text{SYM}} = -\frac{\pi T^2\sqrt{\lambda}}{2} \frac{v}{\sqrt{1 - v^2}},\label{eq22}
\end{equation}
where $\sqrt{\lambda} = \frac{g_{YM}^2 N_c}{4\pi} = \frac{L^2}{\alpha'}$. $\alpha'$ the square of the string length parameter in string theory with unit  $\rm{GeV^{-2}}$. Following In this background, we first need to numerically solve Eq. (\ref{eq18}) to obtain $z_s$, and then use Eq. (\ref{eq21}) to calculate the drag force.

\begin{figure*}
    \centering
    \includegraphics[width=0.9\textwidth]{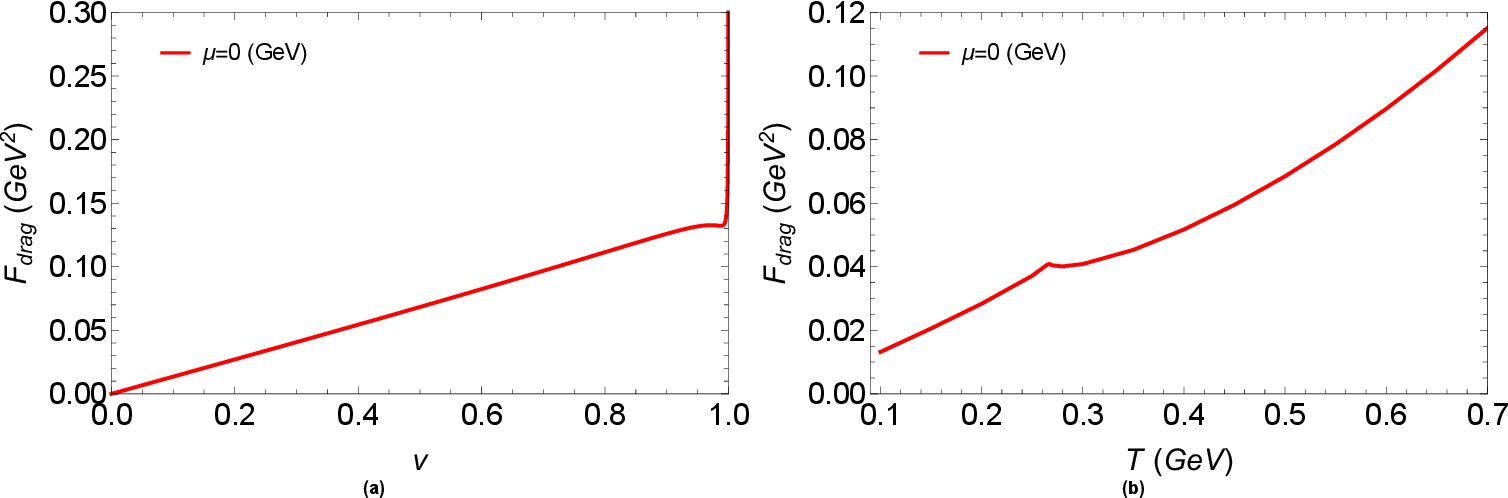}
    \caption{\label{F1} (a) The drag force  as a function of velocity for the pure gluon system at $T = 0.265 \enspace \rm GeV$ and $\mu = 0$. (b) The drag force as a function of temperature for the pure gluon system when a quark is moving at the speed $v = 0.3$ and $\mu = 0$. The unit of $F_{drag}$ is $\rm GeV^2$.
}
\end{figure*}

In Fig. \ref{F1}, we examine the variation of the drag force in a pure gluon system under different conditions. Fig. \ref{F1} (a) shows the relationship between the drag force and velocity in the case of zero chemical potential. As depicted, the drag force escalates continuously with increasing velocity. In Fig. \ref{F1} (b), we present the variation of the drag force along with increasing temperatures under zero chemical potential. The figure reveals an evident increase of the drag force with rising temperature, converging towards its conformal value in the $\mathcal{N}=4$ SYM theory.
Furthermore, the discontinuity of the drag force as a function indicates the first-order phase transition.

\begin{figure*}
    \centering
    \includegraphics[width=0.9\textwidth]{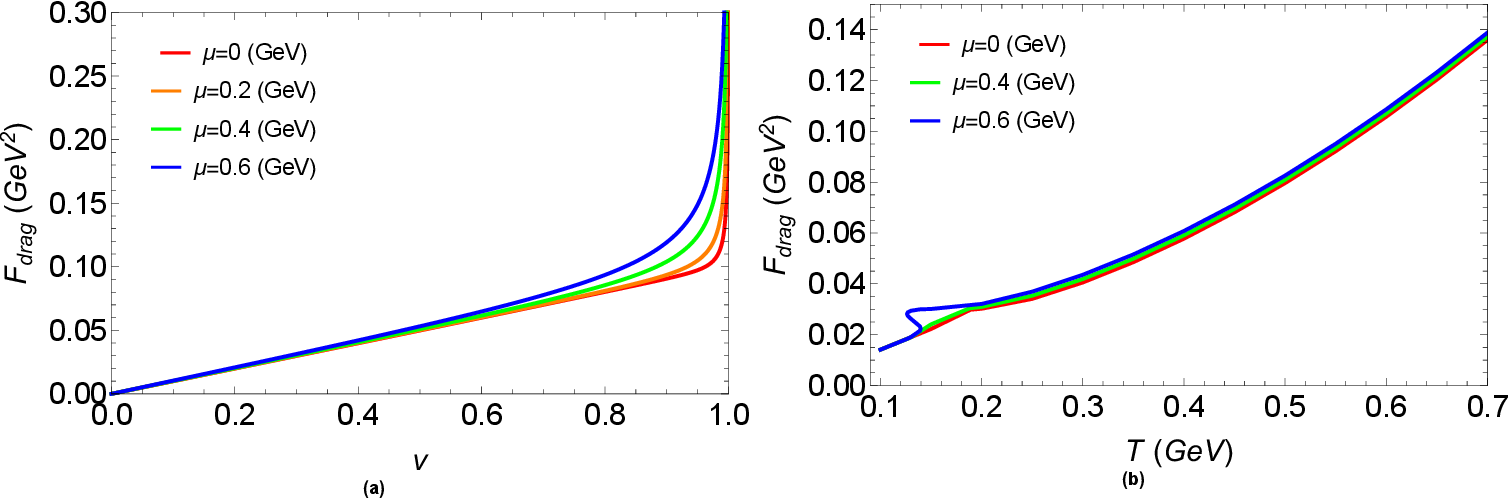}
    \caption{\label{F2}  (a) The drag force as a function of velocity in a 2-flavor system at $T = 0.189 \enspace \rm GeV$ for varying chemical potentials. (b) The drag force as a function of temperature for a 2-flavor system when a quark is moving at the speed $v = 0.3$ under different chemical potentials. The unit of $\mu$ is GeV, and the unit of $F_{drag}$ is $\rm GeV^2$.
}
\end{figure*}

In Fig. \ref{F2}, we can observe the variation of the drag force in a 2-flavor system under different conditions. Fig. \ref{F2} (a) illustrates the drag force as a function of velocity at a fixed temperature for various chemical potentials. It can be seen that as the velocity increases, the drag force also increases, and the rate of this increase becomes more pronounced. Additionally, an increase in the chemical potential further amplifies the rate of increase, meaning that the drag force at a given velocity rises with an increase in chemical potential. Fig. \ref{F2} (b) displays the variation in drag force at different temperatures and chemical potentials in a 2-flavor system, with the quark velocity set at \(v = 0.3\). The graph clearly shows a trend: as the temperature increases, the drag force also continuously increases. This implies that the average kinetic energy of quarks and gluons increases, and thermal motion is enhanced, thus increasing the scattering and interactions faced by a moving quark, leading to a greater drag force. It can also be observed that for larger values of \(\mu\) at the same temperature, the drag force is greater. The chemical potential is generally related to the particle number density, and a larger chemical potential means a higher quark number density in the system. In a denser environment, scattering events among quarks are more frequent, and interactions are stronger, hence a larger drag force. Furthermore, we can see that there is a multiple-value (jump) region for the drag force around the first-order phase transition.

\begin{figure*}
    \centering
    \includegraphics[width=0.9\textwidth]{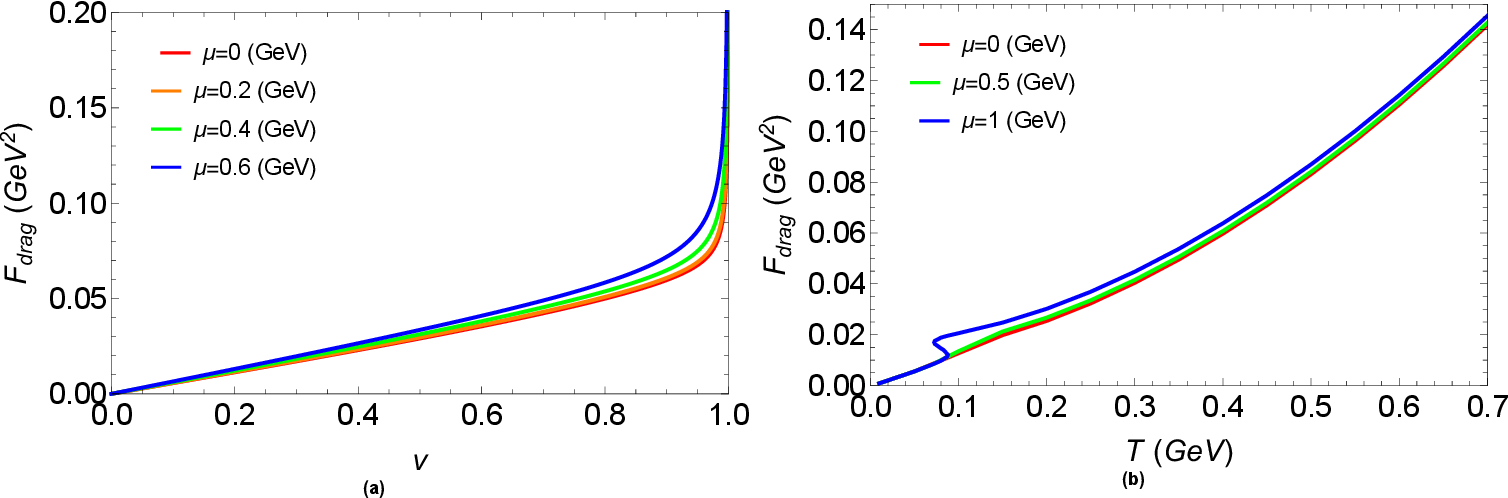}
    \caption{\label{F3}  (a) The drag force as a function of velocity in a 2+1-flavor system at $T = 0.128 \enspace \rm GeV$ for varying chemical potentials. (b) The drag force as a function of temperature for a 2+1-flavor system with a quark moving at the speed $v = 0.3$ under different chemical potentials. The unit of $\mu$ is GeV, and the unit of $F_{drag}$ is $\rm GeV^2$.
}
\end{figure*}

In Fig. \ref{F3}, we observe the variation in the drag force under different conditions within a 2+1 flavor system. Fig. \ref{F3} (a) displays the relationship between the drag force and velocity at various chemical potentials at the critical temperature. It shows that the drag force increases with velocity, indicating that in the 2+1 flavor system, the resistance faced by a moving quark increases with its velocity. The higher the velocity, the greater the medium resistance that the quark needs to overcome. With an increase in chemical potential, the rate at which the drag force increases also becomes more significant. This suggests that at higher chemical potentials, corresponding to a "denser" chemical environment, the quarks experience a more pronounced resistance when increasing their velocity. Fig. \ref{F3} (b) displays the variations in drag force within a 2+1 flavor system under various temperatures and chemical potentials when the velocity of the quark is set at $v = 0.3$. It is clearly shown a trend where the drag force increases as the temperature rises, indicating that the average kinetic energy of quarks and gluons increases, thereby enhancing their thermal motion. As a result, the drag force experienced by a moving quark due to scattering and interaction increases. The impact of chemical potential on drag force is not pronounced at low temperatures but becomes significant at higher temperatures. At a given temperature, a larger \(\mu\) value correlates with a greater drag force; the chemical potential is generally associated with particle number density, and a higher chemical potential implies a denser quark number in the system. In such a dense environment, quarks scatter more frequently, and interactions are stronger, leading to an increased drag force. Additionally, a multiple-value region is observed on the curve at $\mu = 1$, which means that a first-order phase transition has occurred in the system.

\begin{figure*}
    \centering
    \includegraphics[width=0.9\textwidth]{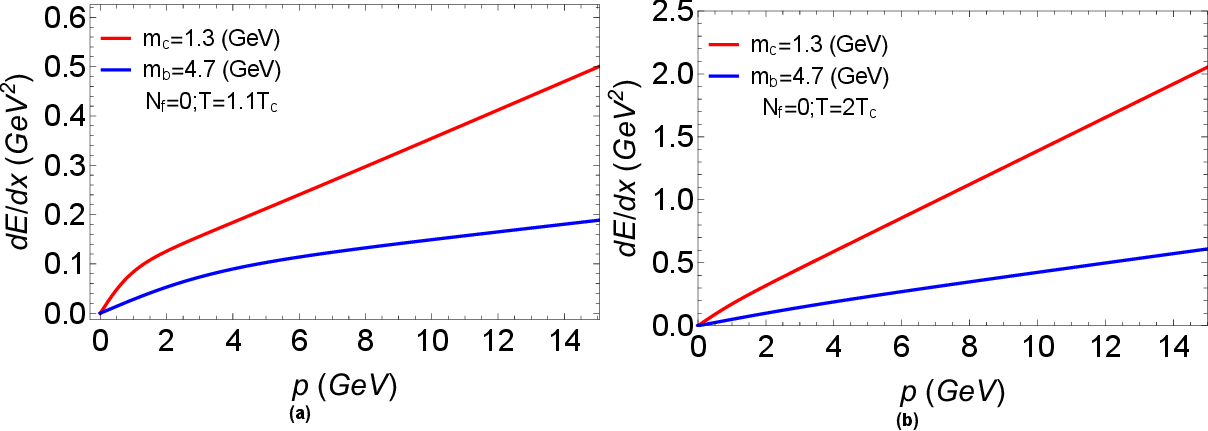}
    \caption{\label{E1}  Energy loss of the bottom ($m_b = 4.7 \enspace \rm GeV$) and charm quark ($m_c = 1.3 \enspace \rm GeV$) in a pure gluon system as a function of momentum $p$($GeV$). (a) The energy loss as a function of momentum at temperature \(T=1.1T_c\). (b) The energy loss as a function of momentum at temperature \(T=2T_c\).
}
\end{figure*}

\begin{figure*}
    \centering
    \includegraphics[width=0.9\textwidth]{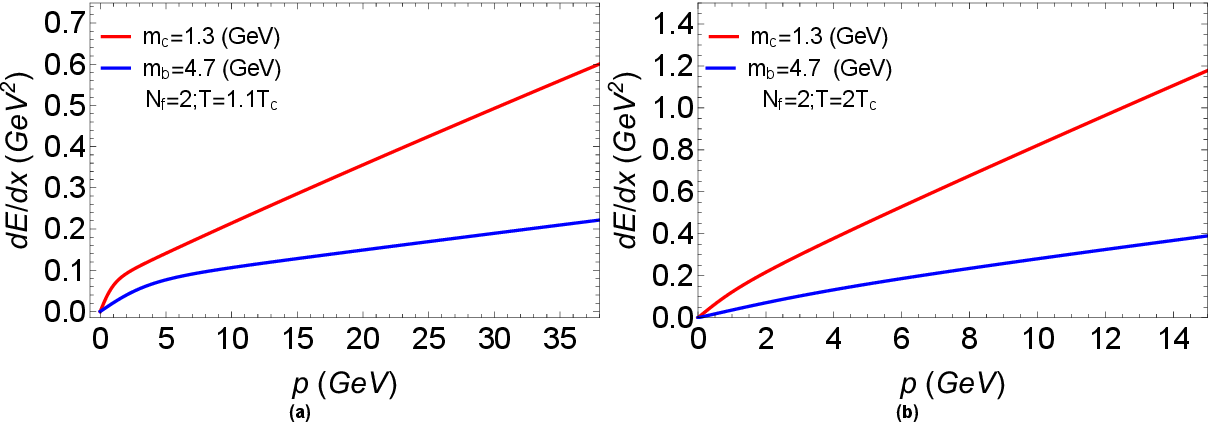}
    \caption{\label{E2}  Energy loss of the bottom ($m_b = 4.7 \enspace \rm GeV$) and charm quark ($m_c = 1.3 \enspace \rm GeV$) in a 2-flavor system as a function of momentum $p$($GeV$). (a) The energy loss as a function of momentum at temperature \(T=1.1T_c\). (b) The energy loss as a function of momentum at temperature \(T=2T_c\).
}
\end{figure*}

\begin{figure*}
    \centering
    \includegraphics[width=0.9\textwidth]{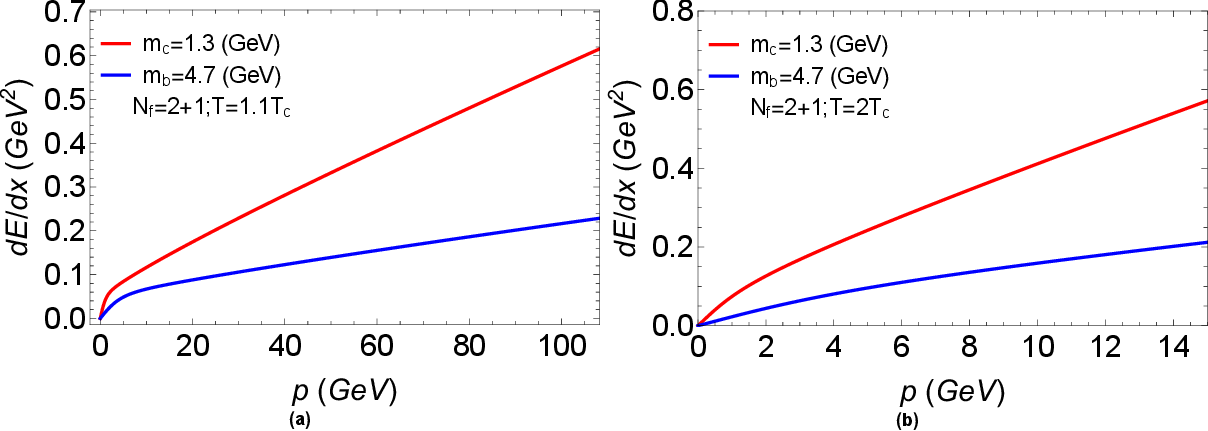}
    \caption{\label{E3}  Energy loss of the bottom ($m_b = 4.7 \enspace \rm GeV$) and charm quark ($m_c = 1.3 \enspace \rm GeV$) in a 2+1-flavor system as a function of momentum $p$ ($GeV$). (a) The energy loss as a function of momentum at temperature \(T=1.1T_c\). (b) The energy loss as a function of momentum at temperature \(T=2T_c\).
}
\end{figure*}

From Eq. (\ref{eq21}), it can be deduced that energy loss is equal to the drag force, allowing us to plot the relationship between energy loss and momentum for different systems. Fig. \ref{E1} illustrates this relationship in a pure gluon system at zero chemical potential for quarks with the bottom ($m_b = 4.7\enspace \rm GeV$) and charm quark ($m_c = 1.3\enspace \rm GeV$) \cite{Guo:2024mgh,Du:2024riq}. Fig. \ref{E2} shows the relationship for a 2-flavor system at zero chemical potential with the bottom ($m_b = 4.7\enspace \rm GeV$) and charm quark ($m_c = 1.3 \enspace \rm GeV$), and Fig. \ref{E3} depicts the same for a 2+1 flavor system at zero chemical potential. From Fig. \ref{E1}, Fig. \ref{E2}, and Fig. \ref{E3}, it is evident that energy loss increases with momentum. The mass of the quarks also affects energy loss; lighter quarks yield greater energy loss. Moreover, the temperature has a more significant impact on energy loss than quark mass, with higher temperatures leading to increased energy loss.

\section{Diffusion Coefficient}
\label{3}
\begin{figure*}
    \centering
    \includegraphics[width=\textwidth]{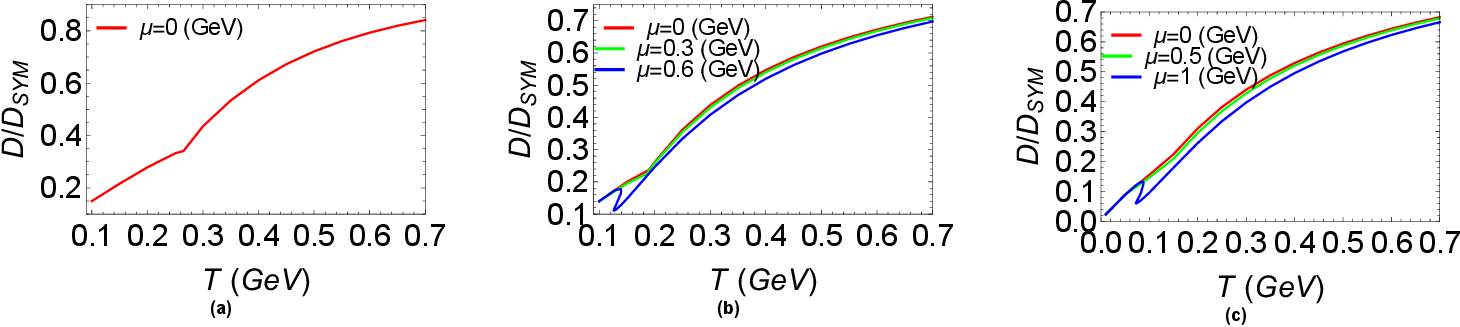}
    \caption{\label{D}  The scaled diffusion coefficient \( D \) /\( D_{SYM} \) as a function of temperature for different systems and different chemical potentials. (a) The pure gluon system, (b) The 2-flavor system, (c) The 2+1 flavor system. The units of temperature and chemical potential are GeV.
}
\end{figure*}

We next proceed with the study of the diffusion coefficient. In the AdS/Schwarzchild background, the drag force Eq. (\ref{eq22}) can be rewritten as  \cite{Gubser:2006qh}
\begin{equation}
F_{\text{drag}}^{\text{SYM}} = -\frac{\pi T^2 \sqrt{\lambda}}{2m} \frac{v m}{\sqrt{1 - v^2}} = -\eta_D p, \label{eq23}
\end{equation}
where \( m \) denotes the mass of the heavy quark, \( \eta_D \) is the drag coefficient, and \( p = \frac{vm}{\sqrt{1 - v^2}} \) is the momentum.
The diffusion time \( t_{\text{SYM}} \) is given by  \cite{Gubser:2006qh}
\begin{equation}
t_{\text{SYM}} = \frac{1}{\eta_D} = \frac{2m}{\pi T^2 \sqrt{\lambda}}, \label{eq24}
\end{equation}
and the diffusion coefficient \( D_{\text{SYM}} \) can be expressed as  \cite{Gubser:2006qh}
\begin{equation}
D_{\text{SYM}} = \frac{T}{m} t_{\text{SYM}} = \frac{2}{\pi T \sqrt{\lambda}}. \label{eq25}
\end{equation}
Eq. (\ref{eq21}) can be rewritten as
\begin{equation}
F_{\text{drag}} = -\frac{e^{2A_s(z_s)}\sqrt{1 - v^2}}{\pi^2 T^2 z_s^2} \frac{\pi T^2 \sqrt{\lambda}}{2m} \frac{vm}{\sqrt{1 - v^2}}, \label{eq26}
\end{equation}
The diffusion time \( t \) is
\begin{equation}
t = \frac{2m}{\pi T^2 \sqrt{\lambda}} \frac{\pi^2 T^2 z_s^2}{e^{2A_s(z_s)} \sqrt{1 - v^2}}. \label{eq27}
\end{equation}
The diffusion coefficient \( D \) can be represented as
\begin{equation}
D = \frac{T}{m} t = \frac{2}{\pi T \sqrt{\lambda}} \frac{\pi^2 T^2 z_s^2}{e^{2A_s(z_s)} \sqrt{1 - v^2}}. \label{eq28}
\end{equation}
From Eq. (\ref{eq25}) and Eq. (\ref{eq28}), we can deduce
\begin{equation}
\frac{D}{D_{\text{SYM}}} = \frac{\pi^2 T^2 z_s^2}{e^{2A_s(z_s)} \sqrt{1 - v^2}}. \label{eq29}
\end{equation}
From Fig. \ref{D}, it can be observed that as the temperature increases, the ratio of the diffusion coefficient to its conformal value in the $\mathcal{N}=4$ SYM theory also increases. This growth indicates that the diffusive properties of the system become more akin to the behaviors predicted by conformal field theory at higher temperatures. The graph also displays that under different chemical potentials, as temperature rises, this ratio ultimately trends towards 1. At lower temperatures, the impact of the chemical potential on the system's properties is more pronounced, but as the temperature escalates, the influence of the chemical potential diminishes, and the role of temperature becomes more dominant. This differential interaction between temperature and chemical potential is key to understanding physical processes such as phase transitions and critical phenomena. The convergence at high temperatures suggests that the diffusion behavior of the system becomes more consistent with predictions from conformal theory, reflecting the conformal behavior that QCD-like theories might exhibit in high-temperature regions. Conversely, at lower temperatures, a larger deviation in this ratio highlights the amplification of non-conformal effects, signifying the emergence of more complex interactions and phenomena when moving away from the high-temperature limit, where QCD-like systems no longer exhibit properties akin to those of simpler conformal theories. Additionally, the graph shows a similar multiple-value region to that observed with the drag force, indicating that when the chemical potential exceeds a certain threshold, the system undergoes the first-order phase transition.

\begin{figure}
    \centering
    \includegraphics[width=\columnwidth]{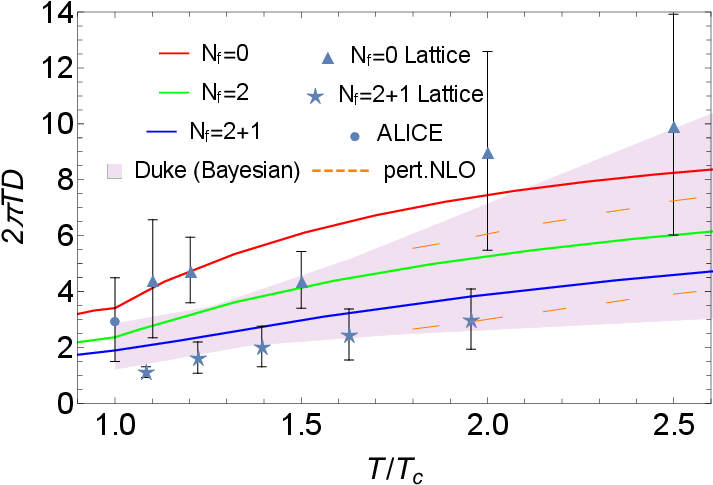}
    \caption{\label{b} The scaled diffusion coefficient $2\pi TD$ as a function of $T/T_c$. The lattice data in the figure are results from Ref. \cite{Altenkort:2023oms}. The results of ALICE are from Ref. \cite{ALICE:2021rxa}. The orange dashed line represents the NLO perturbative calculations \cite{Caron-Huot:2007rwy}.  The pink region represents Bayesian inference results from the Duke hydro/transport model \cite{Xu:2017obm}. The solid curves represent the results of our model.
}
\end{figure}

We compare the spatial heavy quark diffusion coefficient, normalized by $2\pi T$, with the estimates from lattice QCD, ALICE, Next-to-Leading Order (NLO) perturbative predictions, as depicted in Fig. \ref{b}. We can see that the results of our model almost fall within the error bars of the lattice data for $N_f = 0$ and $N_f = 2+1$ \cite{Altenkort:2023oms}. For the 2-flavor system, the curves correspond to our predictive results. Our holographic results for different flavors fit the results of ALICE well \cite{ALICE:2021rxa}. At high temperatures, our results for $D$ show slight discrepancies with the NLO perturbative predictions \cite{Caron-Huot:2007rwy}. Moreover, our holographic results of $N_f = 2$ and $N_f = 2+1$ coincide completely with the region of Duke hydro/transport model \cite{Xu:2017obm}.

\section{Jet Quenching Parameter}
\label{4}
We now turn to the study of the jet quenching parameter. As is well-known, the jet quenching parameter $\hat{q}$ is related to Wilson loops, as shown in the following \cite{Liu:2006ug}
\begin{equation}
\langle W^A[C] \rangle \approx \exp\left(-\frac{1}{4\sqrt{2}} \hat{q} L^{-} L'^2\right), \label{eq30}
\end{equation}
where $W^A[C]$ is the Wilson loop in the adjoint representation, and $C$ is a rectangular contour of size $L' \times L^{-}$. The quark and antiquark are separated by a small $L'$ and travel along the $L^-$.

Simultaneously, the following equation can be used
\begin{equation}
\langle W^A[C] \rangle \approx \langle W^F[C] \rangle^2, \label{eq31}
\end{equation}
and
\begin{equation}
\langle W^F[C] \rangle \approx \exp[-S_I], \label{eq32}
\end{equation}
where $W^F[C]$ is the fundamental representation of the Wilson loop, and $S_I = S - S_0$ (with $S$ being the total energy of the quark-antiquark pair, and $S_0$ the self-energy of isolated quarks and antiquarks).
The general relation for the jet quenching parameter is given by
\begin{equation}
\hat{q} = \frac{8\sqrt{2} S_I}{L^- L'^2}. \label{eq33}
\end{equation}
Using light-cone coordinates \(x^{\pm} = \frac{(x^1 \pm t)}{\sqrt{2}}\), the metric takes the form
\begin{equation}\begin{aligned}
&ds^2 = \frac{L^2 e^{2A_s(z)}}{z^2} [ dx_2^2 + dx_3^2 + \\&\frac{1}{2}(1 - g(z)) \left( dx^{+2} + dx^{-2} \right) + (1 + g(z)) dx^+ dx^- + \frac{dz^2}{g(z)} ]. \end{aligned} \label{eq34}
\end{equation}
With the string parameterized by \( x^\mu(\tau, \sigma) \), the Nambu-Goto action can be written as
\begin{equation}
S_{NG} = \frac{1}{2\pi\alpha'} \int d\sigma d\tau \sqrt{-\det \gamma_{\alpha\beta}}, \label{eq35}
\end{equation}
where \( \gamma_{\alpha\beta} \) is the induced metric on the string worldsheet, with the worldsheet coordinates \( \sigma^\alpha = (\tau, \sigma) \) set as \( (x^-, x_3) \). The selection of a particular static gauge coordinate follows,
\begin{equation}
x^-=\tau,x_3=\sigma,x^+=x_2=const.
\end{equation}
If one assumes a profile where $z$ is a function of $\sigma$, then Eq. (\ref{eq34}) can be rewritten as
\begin{equation}
ds^2 = \frac{L^2 e^{2A_s(z)}}{z^2} \left[ \frac{1}{2}(1 - g(z)) d\tau^2 +(1+ \frac{z'^2}{g(z)})d\sigma^2 \right],
\end{equation}
where $z'=\frac{dz}{d\sigma}$. Thus, \( S_{NG} \) can be written as
\begin{equation}
S_{NG} = \frac{L^-}{\sqrt{2\pi\alpha'}} \int_0^{\frac{L'}{2}} d\sigma \frac{L^2 e^{2A_s(z)}}{z^2} \sqrt{\frac{1 -g(z)}{2} (1+\frac{z'^2}{g(z)})}. \label{eq36}
\end{equation}
Since the Lagrangian density is time-independent, the Hamiltonian of the system is a constant
\begin{equation}
\mathcal{L} - z' \frac{\partial \mathcal{L}}{\partial z'} =\frac{ \Pi_z}{ \sqrt{2}}. \label{eq37}
\end{equation}
From the above equation, \( z' \) is obtained as
\begin{equation}
z' = \sqrt{g(z) \left(\frac{L^4 e^{4A_s(z)}(1 - g(z))}{\Pi_z^2 z^4} - 1 \right)}. \label{eq38}
\end{equation}
Integrating Eq. (\ref{eq38}) results in
\begin{equation}
\frac{L'}{2} = a_0 \Pi_z + O(\Pi_z^3), \label{eq39}
\end{equation}
where \( a_0 \) is defined as
\begin{equation}
a_0 = \int_0^{z_h}dz \, \frac{ z^2 L^{-2}e^{-2A_s(z)}}{ \sqrt{g(z) (1 - g(z))}}. \label{eq40}
\end{equation}
Here, we have considered that for small length \( L' \), the constant \( \Pi_z \) is small, and its higher-order terms can be ignored. Substituting Eq. (\ref{eq38}) into Eq. (\ref{eq36}) yields
\begin{equation}
S_{NG} = \frac{L^-}{ \pi \alpha'} \int_{0}^{z_h} dz \, \frac{L^4 e^{4A_s(z)} (1 - g(z))}{z^2 \sqrt{2g(z) \left(L^4 e^{4A_s(z)}(1 - g(z)) + \Pi_z^2 z^4\right)}}, \label{eq41}
\end{equation}
where we have used \( z' = \frac{\partial z}{\partial \sigma} \). For small \( \Pi_z \), expanding this equation results in:
\begin{equation}\begin{aligned}
&S_{NG} =\frac{L^-}{ \pi \alpha'} \int_{0}^{z_h} dz \, \frac{L^2 e^{2A_s(z)}}{z^2} \sqrt{\frac{1-g(z)}{2g(z)}} (1 + \\&\frac{L^{-4} e^{-4A_s(z)}\Pi_z^2 z^4}{2(1 - g(z))} + \ldots ), \end{aligned}\label{eq42}
\end{equation}
the action diverges and should be subtracted by the self-energy of two separate strings whose worldsheets lie at \( x_2 = \pm \frac{L'}{2} \) and stretch from the boundary to the horizon, as follows
\begin{equation}
S_0 = \frac{L^-}{2 \pi \alpha'} \int_{0}^{z_h} dz \,  \frac{L^2  e^{2A_s(z)}}{z^2} \sqrt{\frac{1-g(z)}{2g(z)}} . \label{eq43}
\end{equation}
Therefore, the normalized action can be written as
\begin{equation}
S_I = S_{NG} - 2S_0 \equiv \frac{L^-\Pi_{z}^2 a_0}{2\sqrt{2}\pi\alpha'}. \label{eq44}
\end{equation}
Inserting Eq. (\ref{eq44}) into Eq. (\ref{eq33}), we obtain the following expression for the jet quenching parameter in the holographic model
\begin{equation}
\hat{q} = \frac{1}{\pi \alpha' a_0}, \label{eq45}
\end{equation}
where we have used Eq. (\ref{eq39}) to represent \( \Pi_z \), and \( a_0 \) is the numerical integral defined in Eq. (\ref{eq40}).
For $\mathcal{N}=4$ supersymmetric Yang-Mills theory, in the large \( N_c \) and large \( \lambda \) limit, Eq. (\ref{eq45}) leads to the following analytic expression \cite{Buchel:2006bv}
\begin{equation}
\hat{q}_{SYM} = \frac{\pi^{3/2} \Gamma\left(\frac{3}{4}\right)}{\Gamma\left(\frac{5}{4}\right)} \sqrt{\lambda}T^3, \label{eq46}
\end{equation}
where \( \Gamma \) denotes the Gamma function.

\begin{figure*}
    \centering
    \includegraphics[width=\textwidth]{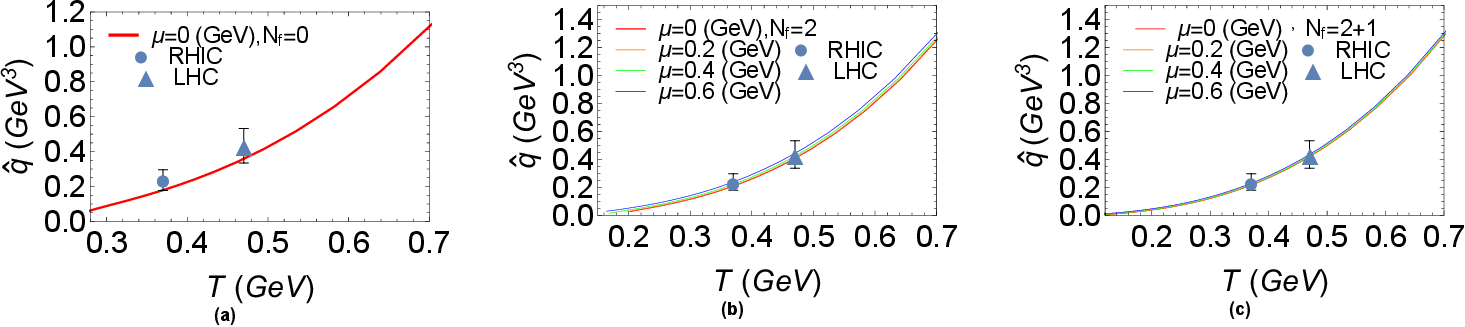}
    \caption{\label{q}   Jet quenching parameter versus temperature for the holographic model with different chemical potentials. Error bars represent experimental values from RHIC and LHC \cite{JET:2013cls}. (a) The pure gluon system, (b) the 2-flavor system, and (c) the 2+1-flavor system. The unit of $\hat{q}$ is $\rm GeV^3$, and the unit of $\mu$ is GeV.
}
\end{figure*}

To obtain the jet quenching parameters in the holographic QCD models for three different systems, we performed numerical solutions for varying values of temperature and chemical potential. The resulting curves are illustrated in the Fig. \ref{q}. It can be observed both the chemical potential and temperature lead to an enhancement of the jet quenching parameter. This indicates that in the model considered, the medium is denser or hotter, resulting in greater energy loss. This is consistent with the physical intuition that jets passing through a medium at higher temperatures or greater chemical potentials (i.e., higher density or more particle-rich environments) will encounter more scattering centers and, hence, experience greater energy loss. Our holographic results for different flavors are consistent with the experimental results of RHIC and LHC \cite{JET:2013cls}.

\begin{figure*}
    \centering
    \includegraphics[width=0.9\textwidth]{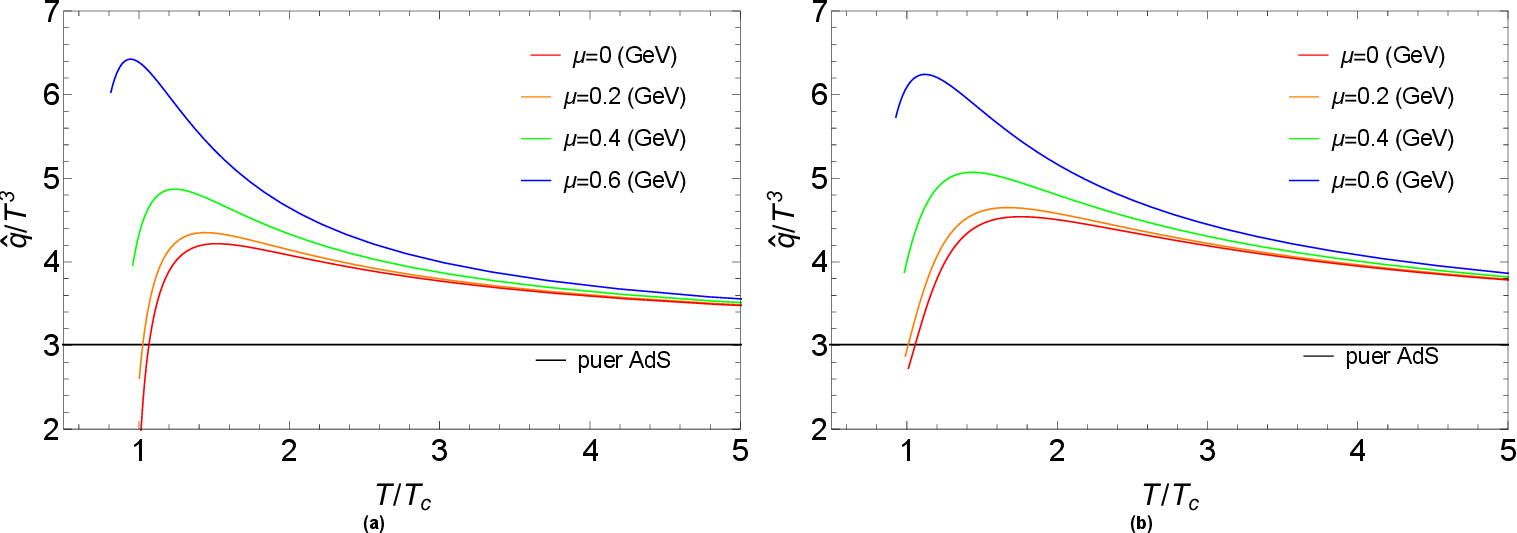}
    \caption{\label{qT} $\hat{q}/T^3$ as a function of temperature $T$. (a) The 2-flavor system, and (b) the 2+1-flavor system. The unit of $\mu$ is in GeV.}
\end{figure*}

\begin{figure}
    \centering
    \includegraphics[width=\columnwidth]{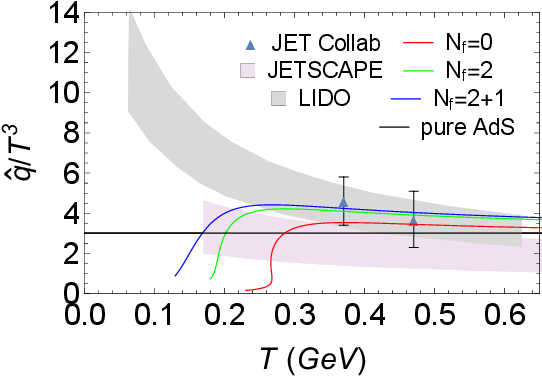}
    \caption{\label{qTT}  The relationship for $\hat{q}/T^3$ at zero chemical potential with temperature $T$, and its comparison with the results taken from Jetscape Collaboration \cite{Soltz:2019aea}, LIDO transport model \cite{Ke:2020clc}, JET Collaboration \cite{JET:2013cls}. The unit of $T$ is GeV.}
\end{figure}

In Fig. \ref{qT}, we have depicted $\hat{q}/T^3$ as a function of temperature, where we observe that the curve exhibits a peak above the phase transition temperature, and subsequently approaches the values corresponding to a pure AdS background. This behavior is markedly different from that obtained in a pure AdS background, where $\hat{q}/T^3$ remains constant across all temperatures. This suggests that dynamic holographic quantum chromodynamics encodes new properties about the deconfining phase transition. In Fig. \ref{qTT}, we compare the graph of $\hat{q}/T^3$ as a function of temperature with other results \cite{Grefa:2023hmf,Apolinario:2022vzg,Ke:2020clc,Cao:2024pxc}. It can be observed that our holographic model fits the Jetscape Collaboration \cite{Soltz:2019aea} well and close to the results of LIDO transport model \cite{Ke:2020clc} at high temperature.

\begin{figure*}
    \centering
    \includegraphics[width=\textwidth]{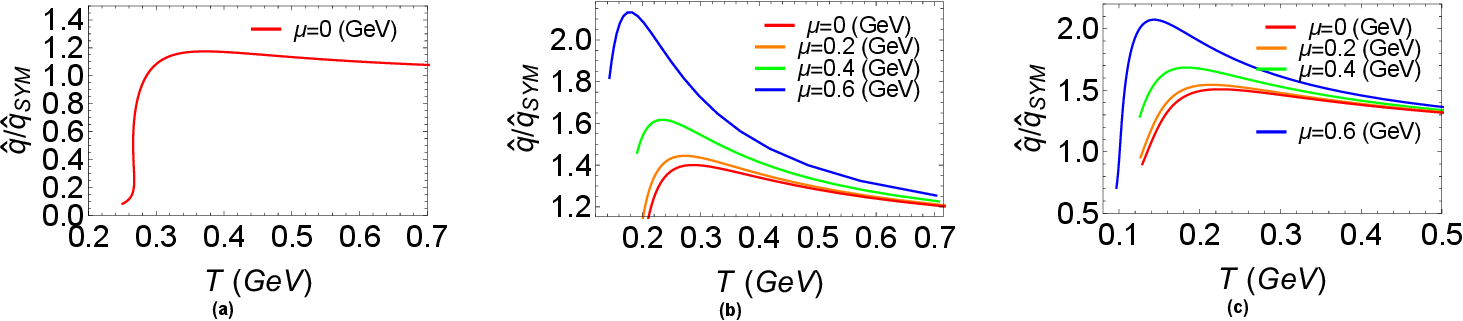}
    \caption{\label{qSYM} The ratio of the jet quenching parameter in the holographic QCD model to \( \hat{q}_{SYM} \). (a) The pure gluon system, (b) the 2-flavor system, (c) the 2+1-flavor system. The unit of $T$ is GeV, and the unit of $\mu$ is GeV.
}
\end{figure*}

In the three systems, according to the temperature calculations at different values of chemical potential \( \mu \), the ratio of the jet quenching parameter in the holographic QCD model to \( \hat{q}_{SYM} \) is seen in Fig. \ref{qSYM}. It can be noticed that at lower temperatures, the jet quenching parameter \( \hat{q} \) is below \( \hat{q}_{SYM} \). This implies that, at lower temperatures, the medium's quenching action on high-energy jets predicted by the holographic QCD model is weaker than the theoretical predictions in the AdS/Schwarzschild background. However, as the temperature increases, the ratio of \( \hat{q} \) to \( \hat{q}_{SYM} \) first grows, indicating an intensification of the quenching effect, then starts to decrease after reaching a certain threshold, the quenching effect is relatively weakened, ultimately approaching 1. This indicates that at higher temperatures, the quenching effect predicted by the holographic QCD model tends to converge with the theoretical predictions made under the AdS/Schwarzschild background.

\section{Conclusion}
\label{5}
One of the scientific goals of heavy-ion collisions is to extract the properties of the medium from combined phenomenological studies of experimental data across various jet quenching measurements \cite{JET:2013cls}. The drag force estimates can be correlated with the measured nuclear modification factor for heavy quarks \cite{Gubser:2009sn}. Transport coefficients are also important inputs to establish the stationary distributions of the transport equations \cite{Walton:1999dy,Bhattacharyya:2024hku}. In this paper, we also show how the jet quenching parameter can inform us about the dynamics of QCD phase transition.

We systematically calculated transport properties, such as drag force, diffusion coefficient, and jet quenching parameter, which are pivotal for understanding the behavior of heavy quarks as they traverse the plasma medium. We did this for systems with $N_f = 0$, $N_f = 2$, and $N_f = 2+1$. These transport properties are being studied for the first time in a holographic model including different quark flavors. We employ the five-dimensional EMD model to investigate the characteristic transport properties of strongly interacting matter. Leveraging a machine-learning-based model, as outlined in Ref. \cite{Chen:2024ckb}, we calculate the drag force, jet quenching parameter, and diffusion coefficient for heavy quark in systems including different quark flavors at finite temperature and chemical potential. The temperature dependence and chemical potential dependence of the diffusion coefficient have significant implications for understanding the transport dynamics of the QGP in heavy-ion collision experiments.

A detailed examination of the drag force demonstrates how temperature, velocity, and chemical potential interplay to affect the magnitude and trend of the force experienced by quarks. An increase in either temperature or velocity leads to a rise in the drag force. Moreover, a higher chemical potential further intensifies this effect in the 2-flavor and 2+1-flavor systems. Analysis of the diffusion coefficients in our holographic calculation unveils the temperature and chemical potential dependence of quark diffusion in the strongly coupled plasma. As temperature increases, the ratio of the diffusion coefficient to its conformal value in the \(\mathcal{N}=4\) SYM theory also increases. This suggests that at higher temperatures, the QGP's diffusion properties approach those of conformal field theories, a behavior naturally required by our model. The jet quenching parameter, like other holographic models \cite{Grefa:2022sav,Li:2014hja,Cao:2024jgt,Zhou:2022izh}, shows a peak above the phase transition temperature. The presence of dynamic quarks reduces the drag force at fixed velocity; specifically, the drag force decreases with the number of flavors in the system. At fixed momentum, the energy loss increases as the number of flavors grows. Furthermore, the dimensionless ratio $\frac{\hat{q}}{T^3}$ shows a monotonic rise with temperature $T$ for systems containing more flavors. Finally, the diffusion coefficient decreases with increasing flavor number, demonstrating a trend consistent with lattice QCD simulations.

The results validates the model's predictive capability of our model presented in \cite{Chen:2024ckb} and this paper offers a comprehensive numerical and theoretical analysis of jet quenching mechanisms within the QGP, a key phenomenon for understanding energy loss during jet propagation in heavy-ion collisions. Our results are consistent with the experiments and other theoretical models, which confirm the validity and robustness of our EMD model. Our studies can offer useful insights into the non-perturbative aspects of QCD. is worth mentioning that our model is expected to be mean-field results for the critical exponents. How to obtain results beyond the mean-field level in holographic QCD models is an interesting issue, which we hope to address in future work.

As a outlook, we can incorporate the error bars from the lattice QCD results into models such as the Bayesian holographic model. This work serves as the foundation for our future work on the Bayesian holographic model. Additionally, we will explore the transport properties of the Bayesian holographic model in the future.

\vskip 0.5cm
{\bf Acknowledgement}
\vskip 0.2cm
This work is supported in part by the National Natural Science Foundation of China (NSFC) Grants No. 12175100,12405154, the CUHK-Shenzhen university development fund under grant No. UDF01003041 and the BMBF funded KISS consortium (05D23RI1) in the ErUM-Data action plan.

\bibliographystyle{unsrtnat}
\bibliography{ref}
\end{document}